\documentstyle[emulateapj]{article}

\lefthead{Alonso-Herrero \& Knapen}
\righthead{Circumnuclear H\,{\sc ii} regions in nearby galaxies}

\begin{document}

\title{Statistical
properties of circumnuclear H\,{\sc ii} regions in nearby
galaxies\footnote{Based on observations with the NASA/ESA Hubble Space
Telescope, obtained from the data Archive at the Space Telescope Science
Institute, which is operated by the Association of Universities for
Research in Astronomy, Inc., under NASA contract NAS 5-26555. }}

\author{Almudena Alonso-Herrero}
\affil{University of Hertfordshire, Department 
of Physical Sciences, Hatfield, Herts AL10 9AB, UK}
\affil{Present address: Steward Observatory, The University of Arizona, Tucson, AZ 85721, USA}
\and
\author{Johan H. Knapen}
\affil{Isaac Newton Group of Telescopes, Apartado 321, 
        E-38700 Santa Cruz de la Palma, Tenerife, Spain}
\affil{and University of Hertfordshire, Department 
of Physical Sciences, Hatfield, Herts AL10 9AB, UK}

\begin{abstract}

We analyze the statistical properties of the circumnuclear H\,{\sc ii}
regions of a sample of 52 nearby galaxies ($v < 1000\,{\rm
km}\,{\rm s}^{-1}$) from archival {\it HST}/NICMOS $H$-band and
Pa$\alpha$ ($1.87\,\mu$m) observations at unprecedented spatial
resolutions of between 1 and 30\,pc. We catalog H\,{\sc ii} regions from
the continuum-subtracted Pa$\alpha$ images, and find  H\,{\sc
ii} regions in the 
central regions of most galaxies, and more than a hundred in 8 galaxies. In
contrast to disk H\,{\sc ii} regions, the physical properties 
(luminosity and size) of
individual circumnuclear H\,{\sc ii} regions do not vary strongly with the
morphological type of the host galaxy, nor does the number of
circumnuclear H\,{\sc ii} regions per unit area. The H$\alpha$
luminosity within the central kpc, as derived from H\,{\sc ii} region 
emission, is significantly enhanced in early-type (S0/a--Sb) galaxies.
We find evidence that bars increase the circumnuclear star formation,
presumably by funneling gas from the disk towards the nucleus. 
Barred galaxies exhibit 
enhanced luminosities of the brightest H\,{\sc ii} region, 
the central kpc 
H$\alpha$ luminosities (an effect mostly due to the early type galaxies
in our sample), and the star formation rates per unit stellar
mass (which could also be understood as 
the integral equivalent widths of Pa$\alpha$) 
over the central kpc with respect to non-barred galaxies.

We fit the luminosity functions and 
diameter distributions of the circumnuclear H\,{\sc ii} regions
in 8 galaxies where we can catalog enough H\,{\sc ii} regions to do so
in a meaningful way. We use power laws and find that the
 fitted slopes of the H\,{\sc ii} region LF are exactly in the
previously found ranges, and even confirm a trend with
steeper slopes in galaxies of earlier morphological type. This 
implies that the physical processes giving rise to enhanced star
formation in the circumnuclear regions of galaxies 
must be similar to those in disks.

\end{abstract}


\section{Introduction}

Extragalactic H\,{\sc ii} regions provide important clues to the physics of
the star formation (SF) process because they directly trace the most
massive stellar populations, and are readily observable through hydrogen
recombination lines (mainly H$\alpha$). Ground-based
H$\alpha$ narrow-band observations of galaxies are easily obtained and
allow the study of the statistics of large numbers of extragalactic
H\,{\sc ii} regions (see Kennicutt 1998a for a review on the advantages
and limitations of the H$\alpha$ emission line for the study of SF in
galaxies). H\,{\sc ii} region populations in galaxies have been studied
for over 30 years now (see early review by Hodge 1974), and have been
based on ever improving observational material -- from photographic
plates (e.g., Hodge 1969; Kennicutt, Edgar \& Hodge 1989, hereafter
KEH), via ground-based CCD imaging on 1-m and 2-m (e.g., Rand 1992;
Banfi et al. 1993; Feinstein 1997) to 4-m class telescopes (e.g., Cepa
\& Beckman 1989, 1990; Knapen et al. 1993; Rozas, Beckman \& Knapen
1996a; Wyder, Hodge, \& Skelton 1997; Knapen 1998), to, most recently,
imaging with the Hubble Space Telescope ({\it HST}; Pleuss, Heller, \&
Fricke 2000; Scoville et al.  2001; this paper).

The main results from statistical studies of more or less complete
ensembles of H\,{\sc ii} regions in disks of galaxies are integral
diameter distributions, and luminosity functions (LFs). The latter are
usually fitted by a power-law of the form:

\begin{equation}
{\rm d}N(L) = A\,L^\alpha\,{\rm d}L
\end{equation}

\noindent where the index $\alpha$ describes the slope of the LF, and
has values of $\alpha=-2.0\pm0.5$ for extragalactic H\,{\sc ii} regions
(e.g., KEH). A similar index has been found for the LF of Galactic radio
H\,{\sc ii} regions (McKee \& Williams 1997, and references therein).

The most interesting features in observed H\,{\sc ii} region LFs are (1)
a change in slope, with a steeper slope at the high-$L$ end of the LF
(e.g., KEH; Rand 1992; Rozas et al. 1996a; Knapen 1998), which has been
explained in terms of differences in the molecular gas cloud spectrum
(Rand 1992), in terms of a transition from single ionizing stars to
ionizing clusters (McKee \& Williams 1997; Oey \& Clarke 1998), in terms
of a transition from ionization to density bounded H\,{\sc ii} regions
(Beckman et al. 2000), and in terms of blending introduced by
observations at too low spatial resolution (Pleuss et al. 2000); (2)
steeper slopes in the interarm regions of some spiral galaxies as
compared to the arm regions, observed in some galaxies (KEH; Rand 1992;
Banfi et al. 1993; Thilker, Braun \& Walterbos 2000; Scoville et
al. 2001), but not in many others (Knapen et al. 1993; Rozas et
al. 1996a; Knapen 1998); (3) a correlation between the LF slopes and the
Hubble morphological type, with the slopes being systematically steeper
in early-type galaxies (KEH; Banfi et al. 1993); and (4) a local maximum
in the LFs around $\log L({\rm H}\alpha) = 
38.6 \pm 0.1$ erg\,s$^{-1}$, interpreted as
evidence for a change from an ionization to a density bounded regime of
H\,{\sc ii} regions of increasing luminosity by Beckman et al. (2000),
but as an artifact of low resolution imaging by Pleuss et al. (2000).

From the theoretical point of view, the observational differences of
H\,{\sc ii} region LFs among galaxies have been reproduced in terms of
evolutionary effects (e.g., different ages, and SF histories), and
properties of the ionizing clusters such as different initial mass
functions (IMFs) and the maximum number of ionizing stars per cluster
(e.g., von Hippel \& Bothun 1990; Feinstein 1997; McKee \& Williams
1997; Oey \& Clark 1997, 1998; Beckman et al. 2000; Scoville et
al. 2001).  It is  generally assumed that the shape and properties of the
H\,{\sc ii} region LF directly 
reflect certain characteristics of the most recent
SF history of the galaxy.

So far, all statistical studies of the properties of extragalactic
H\,{\sc ii} regions have been performed on samples of disk H\,{\sc ii}
regions, whereas the properties of nuclear and circumnuclear H\,{\sc ii}
regions are relatively unknown, due to the combined effects of limited
spatial resolution and blending. The central, kiloparsec scale, regions
of disk galaxies, however, are interesting because of the usually
increased circumnuclear or nuclear SF activity there, and because the
central regions act as excellent laboratories where, e.g., SF and gas
flow processes in barred and non-barred galaxies can be studied in
detail (e.g., review by Knapen 1999). Even though gas fractions and star
formation rates (SFRs) in the central regions can be significantly
enhanced, there is as yet no convincing evidence that the SF processes
in these regions are fundamentally different from those 
occurring in the disks of
galaxies, nor for a different IMF in the central regions of galaxies, or
in fact in general (e.g., reviews by Elmegreen 1998; Gilmore 2001, but
see Eisenhauer 2001 for a differing view on the IMF).

We have used {\it HST}/NICMOS Pa$\alpha$ images of the circumnuclear
regions of a sample of 52 spiral and irregular galaxies in order to
study the statistical properties of H{\sc ii} regions at high spatial
resolution ($1-30\,$pc).  We describe the sample, data, and production
of the H{\sc ii} region catalogs in Section~2, and the statistical
properties of the H{\sc ii} regions in Section~3 and 4. We discuss our
results and summarize our conclusions in Section~5.

\begin{deluxetable}{cllll}
\tablewidth{11cm}
\tablecaption{The sample.}
\tablehead{\colhead{Galaxy}  & \colhead{Hubble} & \colhead{Morphological}  
& \colhead{$v$} & \colhead{Tully} \\
        & \colhead{type}   & \colhead{type}  &  
\colhead{(km s$^{-1}$)} & \colhead{distance}}
\startdata

NGC~247  & 7  & SAB(s)d   & 159  & 2.1 \\
NGC~598  & 6  & SA(s)cd   & $-182$ & 0.7\\
NGC~628  & 5  & SA(s)c    & 659  & 9.7 \\
NGC~672  & 6  & SB(s)cd   & 425  & 7.5 \\
NGC~891  & 3  & SA(s)b?   & 529  & 9.6 \\
NGC~925  & 7  & SB(s)c    & 554  & 9.4\\
NGC~2366 & 10 & IB(s)m    & 102  & 2.9 \\
NGC~2403 & 6  & SAB(s)cd  & 132  & 4.2 \\
NGC~2681 & 0  & (R')SAB(rs)0/a & 710  & 13.3 \\
NGC~2683 & 3  & SA(rs)b   & 415  & 5.7 \\
NGC~2841 & 3  & SA(r)b    & 637  & 12.0 \\
NGC~2903 & 4  & SB(s)d    & 554  & 6.3 \\
NGC~2976 & 5  & SAcpec    & 9    & 2.1 \\
NGC~3077 & 13 & I0pec     & 10   & 2.1\\
NGC~3184 & 6  & SAB(rs)cd & 599  & 8.7 \\
NGC~3593 & 0  & SA(s)0/a  & 628  & 5.5 \\
NGC~3627 & 3  & SAB(s)b   & 737  & 6.6 \\
NGC~3675 & 3  & SA(s)b    & 771  & 12.8 \\
NGC~3738 & 10 & Irr       & 225  & 4.3\\
NGC~3769 & 3  & SB(r)b    & 714  & 17.0\\
NGC~3782 & 6  & SAB(s)cd      & 740  & 17.0\\
IC~749   & 6  & SAB(rs)cd & 784  & 17.0\\
IC~750   & 2  & Sab       & 713  & 17.0 \\
NGC~4062 & 5  & SA(s)c    & 769  & 9.7 \\
NGC~4085 & 5  & SAB(s)c   & 714  & 17.0\\
NGC~4102 & 3  & SAB(s)b   & 862  & 17.0\\
NGC~4136 & 5  & SAB(s)c   & 618  & 9.7 \\
NGC~4144 & 6  & SAB(s)cd  & 267  & 4.1\\        
NGC~4178 & 8  & SB(rs)dm  & 381  & 16.8\\
NGC~4183 & 6  & SA(s)cd   & 934  & 17.0\\
NGC~4190 & 10 & Impec     & 234  & 2.8\\
NGC~4192 & 2  & SAB(s)ab  & $-142$ & 16.8 \\
NGC~4293 & 0  & (R)SB(s)0/a & 882  & 17.0\\
NGC~4294 & 6  & SB(rs)cd  & 415  & 16.8 \\
NGC~4299 & 7  & SAB(s)dm  & 238  & 16.8 \\
NGC~4389 & 6  & SB(rs)bc  & 718  & 17.0 \\
NGC~4395 & 9  & SAd       & 318  & 3.6 \\
NGC~4449 & 10 & IBm       & 200  & 3.0\\
NGC~4559 & 6  & SAB(rs)cd & 816  & 9.7 \\
NGC~4571 & 6  & SA(r)d    & 348  & 16.8 \\
NGC~4605 & 5  & SB(s)c pec & 140  & 4.0\\
NGC~4701 & 6  & SA(s)cd   & 727  & 20.5\\
NGC~4826 & 2  & (R)SA(rs)ab & 414  & 4.1 \\
NGC~5055 & 4  & SA(rs)bc    & 497  & 7.2\\
NGC~5474 & 6  & SA(s)cd    & 277  & 6.0 \\
NGC~5585 & 7  & SAB(s)d    & 303  & 7.0\\
IC~5052  & 7  & SBd        & 598  & 6.7 \\
NGC~6207 & 5  & SA(s)c     & 852  & 17.4 \\
IC~4710  & 9  & SB(s)m     & 741  & 8.9\\
NGC~6744 & 4  & SAB(r)bc   & 842  & 10.4\\
NGC~6822 & 10 & Im         & $-56$  & 0.7\\
NGC~6946 & 6  & SAB(rs)cd  & 46   & 5.5 \\
\enddata
\end{deluxetable}

\section{The data}
\subsection{The sample}

We have selected a sample of galaxies from the {\it HST}/NICMOS Pa$\alpha$
snapshot survey of nearby galaxies as published by B\"oker et al. (1999),
imposing the criteria that galaxies have morphological type S0/a or later,
and velocities of $v < 1000\,{\rm
km\,s}^{-1}$. The B\"oker et al. (1999) ``snapshot'' survey consists of
galaxies selected at random from a master list taken from the Revised Shapley
Ames Catalog (RSA) according to {\it HST} scheduling convenience. Our sample
contains a total of 52 galaxies, and is presented in Table~1, which lists the
Hubble type, heliocentric velocity 
from the Nearby Galaxies Catalog (Tully 1988),  
morphological type, and the distance from the Tully
(1988) catalog assuming $H_0 = 75\,{\rm km\,s}^{-1}\,{\rm Mpc}^{-1}$.

In terms of the morphological type, 25\% of the sample are galaxies
with $T = 0 - 3 $ (that is, morphological
types earlier than Sbc), 19\% are galaxies with $T =4-5$ (Sbc--Sc) 
and 56\% galaxies with types $T > 5$ (morphological types
later than Sc). According to the presence or absence of 
a bar the sample is divided into: 40\% unbarred (A), 25\% strongly 
barred (B) and 35\% weakly barred (AB). 

We determined whether there are distance biases within these 
subclasses, and found that the weakly barred galaxies are at slightly
larger distances: galaxies with type AB are at a median distance of 
9.7\,Mpc, unbarred galaxies at 5.6\,Mpc, and strongly barred galaxies 
at 7.5\,Mpc. We also found that the early-type galaxies in our sample are more
distant: $T=0-3$ are at a median distance of 14.5\,Mpc,
whereas $T=4-5$ and $T >5$ are at median distances of 
9.5\,Mpc and 7.5\,Mpc, respectively. These biases are accounted for when 
we analyze
properties dependent on the distance, as detailed below.

\subsection{Observations and Data Reduction}
Images were taken with the NIC3 camera (pixel size 0.2\arcsec \
pixel$^{-1}$) of NICMOS on the {\it HST} using the broad-band F160W
filter (equivalent to a ground-based $H$-band filter), and the
narrow-band F187N filter. The latter filter ($\Delta \lambda/\lambda
\simeq 1\%$) contains the emission line of Pa$\alpha$ and the adjacent
continuum at $1.87\,\mu$m.  The field of view of the images is
$51.2\arcsec \times 51.2\arcsec$, which in our galaxies corresponds to
the central $175\,{\rm pc}- 5\,{\rm kpc}$, depending upon the distance
to the galaxy.

The images were reduced following standard procedures (see
Alonso-Herrero et al. 2000 for more details). The angular resolution
(FWHM) of the NIC3 images is 0.3\arcsec, as measured from the point
spread function (PSF) of stars in the images.  This corresponds to
typical spatial resolutions of between 1\,pc and 30\,pc for the galaxies
in our sample, using the distances given in Table~1.

The flux calibration of the F160W and F187N images was performed using
conversion factors based upon measurements of the standard star P330-E,
taken during the Servicing Mission Observatory Verification (SMOV)
program (M. J. Rieke, private communication 1999). Unfortunately, there
are no observations available of the continuum adjacent to Pa$\alpha$,
so we used the flux calibrated images at $1.6\,\mu$m as continuum for
the F187N images. As discussed in Maiolino et al. (2000), ideally
narrow-band continuum images on both sides of the emission line are needed to
perform an accurate continuum subtraction which takes into account the
spatial distribution of the extinction. We fine-tuned the scaling
 of the continuum image so that the continuum subtracted
Pa$\alpha$ image (flux calibrated) did not show negative values. This
method generally produces satisfactory results, except in those cases
where the differential extinction between $1.60\,\mu$m and $1.87\,\mu$m
is significant (typically edge-on galaxies and the nuclei of some
galaxies). As we shall see in the next section, the H\,{\sc ii} region
photometry software takes into account the local background, so the
continuum subtraction is not a dominant source of error, although 
errors associated with the background subtraction depend on the
luminosity of the H\,{\sc ii} region.

Throughout the paper we have converted the Pa$\alpha$ luminosity 
into the more commonly used H$\alpha$ luminosity, assuming case B 
recombination ($\frac{{\rm
H}\alpha}{{\rm Pa}\alpha} = 8.7$).
We also stress that the effects of reddening in the central regions will
be attenuated with the use of Pa$\alpha$ luminosities with 
respect to H$\alpha$ luminosities ($A({\rm
Pa}\alpha) \simeq 0.2\,A({\rm H}\alpha$); Rieke \& Lebofsky 1985).

\subsection{Production of the H\,{\sc ii} region catalogs}

The H\,{\sc ii} region catalogs were produced using the software {\sc
region}, kindly provided by Dr. C. H. Heller (see Pleuss et al. 2000,
and references therein for a detailed description).  {\sc region} is a
semi-automated method to locate and compute statistics of H\,{\sc ii}
regions in an image, based on contouring, and taking into account the
local background. The lower limit for the size of an H\,{\sc ii} region
is set to 9 contiguous pixels, which corresponds to {\it minimum} linear
sizes of between 2 and 60\,pc. Each pixel must have an intensity above
the local background of at
least three times the rms noise of that local background (see Rand 1992
and Knapen et al. 1993 for more details on the criteria employed). After
identifying the H\,{\sc ii} regions, the program measures their
position, size (area) and luminosity by subtracting the closest local
background from the observed flux. For those galaxies with detected
H\,{\sc ii} regions, in Table~2 we list 
the number of H\,{\sc ii} regions identified, the H$\alpha$ luminosity
of the brightest, faintest, and median H\,{\sc ii}, as well as 
the diameter of the largest and median H\,{\sc ii} region.
A total of 10  galaxies were excluded from most
of the H\,{\sc ii} region analyses for various reasons, as detailed in the
footnote to Table~2

There are two methods to define the extent of an H\,{\sc ii} region:
percentage-of-peak photometry (PPP) and fixed-threshold photometry
(FTP). The former defines the area to be assigned to an H\,{\sc ii}
region within an isophote whose brightness is a fixed percentage of the
intensity peak, and avoids problems of S/N at the boundary of the
region.  The latter uses a limiting  surface brightness
method to define an H\,{\sc ii} region (see Kingsburgh \& McCall 1998
for a detailed comparison of the two methods).

We used the FTP method, but we still needed to impose the
condition of a minimum size (in pixels) for identifying an H\,{\sc ii}
region. Thus for galaxies at different distances this corresponds to
differing {\it minimum} physical sizes. This effect can be clearly seen
in the median diameters (in pc) of the detected H\,{\sc ii} regions
found for each galaxy (Table~2). In order to show the effect of distance
on the derived properties, the galaxies in Table~2 are sorted by
increasing distance.

\begin{deluxetable}{clllllllll}
\small
\tablewidth{17.5cm}
\tablecaption{Statistical properties of circumnuclear  H\,{\sc ii} regions
in galaxies.}
\tablehead{\colhead{Galaxy}  & \colhead{Dist} & 
\colhead{no} & \colhead{$\log L_{\rm br}({\rm H}\alpha$)} & 
\colhead{$\log L_{\rm f}({\rm H}\alpha$)} & 
\colhead{$\log L_{\rm med}({\rm H}\alpha$)} &
\colhead{diam$_{\rm med}$} & \colhead{diam$_{\rm lar}$} &
\colhead{$\log L_{\rm kpc}({\rm H}\alpha$)} & \colhead{$M_{\rm F160W}$}\\
        & (Mpc)    & H\,{\sc ii}
& (erg s$^{-1}$) & (erg s$^{-1}$) & (erg s$^{-1}$)
& (pc) & (pc) & (erg s$^{-1}$)\\
(1) & (2) & (3) & (4) & (5) & (6) & (7) & (8) & (9) & (10)}

\startdata

NGC~598  &  0.7 &  65 & 36.74 & 35.28 & 35.62 &   3 &   9 &  \nodata 
& \nodata \\
NGC~6822 &  0.7 &   4 & 35.98 & 35.64 & 35.70 &   3 &   4 &  \nodata 
& \nodata \\
NGC~2976 &  2.1 &  17 & 37.47 & 36.17 & 36.62 &  11 &  24 &  \nodata 
& \nodata \\
NGC~3077 &  2.1 & 155 & 38.30 & 36.12 & 36.79 &  10 &  28 &  \nodata 
& \nodata \\
NGC~4190 &  2.8 &   5 & 36.78 & 36.17 & 36.55 &  13 &  16 &  \nodata 
& \nodata \\
NGC~2366 &  2.9 &   3 & 36.70 & 36.44 & 36.62 &  12 &  13 & \nodata 
& \nodata \\\
NGC~4449 &  3.0 & 155 & 38.32 & 36.39 & 36.90 &  14 &  36 & 39.50 & -18.39\\
NGC~4605 &  4.0 &  40 & 37.92 & 36.50 & 36.98 &  17 &  47 & 38.84 & -18.88\\
NGC~4144 &  4.1 &   7 & 38.17 & 37.05 & 37.30 &  14 &  31 & 38.43 & -17.77\\
NGC~4826 &  4.1 & 184 & 38.34 & 36.86 & 37.26 &  16 &  45 & 39.72 & -20.95\\
NGC~2403 &  4.2 &  39 & 37.98 & 36.69 & 36.92 &  17 &  46 & 38.71 & -19.55\\
NGC~3738 &  4.3 &  52 & 38.35 & 36.50 & 37.05 &  21 &  72 & 39.10 & -18.08\\
NGC~3593 &  5.5 & 103 & 38.77 & 37.09 & 37.81 &  26 &  59 & 40.05 & -19.98\\
NGC~6946 &  5.5 & 252 & 39.80 & 36.84 & 37.39 &  25 &  93 & 40.33 & -19.45\\
NGC~5474 &  6.0 &  12 & 37.62 & 36.89 & 37.09 &  22 &  35 & 38.25 & -18.00\\
NGC~2903 &  6.3 & 111 & 39.42 & 37.02 & 38.35 &  28 & 101 & 40.34 & -20.58\\
NGC~3627$^*$ &  6.6 &  28 & 38.47 & 36.24 & 37.67 &  26 &  48 & 39.69 & -21.02\\
IC~5052  &  6.7 &  14 & 39.24 & 37.39 & 37.94 &  36 &  97 & 39.52 & -17.51\\
NGC~5585 &  7.0 &   8 & 37.80 & 36.98 & 37.50 &  29 &  42 & 38.08 & -17.88\\
NGC~5055 &  7.2 & 100 & 38.24 & 37.22 & 37.54 &  27 &  51 & 39.32 & -20.59\\
NGC~672  &  7.5 &  34 & 38.47 & 36.98 & 37.26 &  30 &  72 & 39.01 & -18.26\\
NGC~3184$^*$ &  8.7 &  19 & 38.77 & 37.22 & 37.70 &  39 &  69 & 39.27 & -18.15\\
NGC~925  &  9.4 &  17 & 38.89 & 37.46 & 37.81 &  41 &  78 & 38.50 & -18.45\\
NGC~628  &  9.7 &  13 & 38.30 & 37.54 & 37.80 &  38 &  52 & 38.77 & -19.77\\
NGC~4062 &  9.7 &  25 & 38.02 & 37.28 & 37.55 &  36 &  59 & 38.39 & -19.22\\
NGC~4559 &  9.7 &  20 & 38.89 & 37.89 & 38.09 &  38 &  69 & 39.20 & -19.10\\
NGC~4136 &  9.7 &  23 & 38.19 & 37.12 & 37.32 &  35 &  55 & 38.29 & -18.26\\
NGC~3675 & 12.8 &  68 & 38.51 & 37.57 & 37.86 &  49 &  89 & 40.03 & -21.12\\
NGC~4178 & 16.8 &  24 & 39.34 & 37.86 & 38.32 &  70 & 206 & 39.36 & -18.62\\
NGC~4192 & 16.8 &  44 & 39.52 & 37.88 & 38.52 &  76 & 158 & 40.36 & -21.44\\
NGC~4294 & 16.8 &  60 & 39.34 & 37.85 & 38.30 &  76 & 160 & 39.38 & -18.59\\
NGC~4299 & 16.8 &  39 & 39.52 & 37.83 & 38.34 &  74 & 192 & 39.44 & -17.77\\
NGC~3769 & 17.0 &  35 & 39.17 & 37.84 & 38.14 &  67 & 157 & 39.74 & -19.56\\
NGC~3782 & 17.0 &  25 & 39.14 & 37.72 & 38.17 &  74 & 145 & 38.27 & -18.32\\
IC~749   & 17.0 &  20 & 38.61 & 37.75 & 37.98 &  66 & 115 & 38.49 & -18.21\\
IC~750   & 17.0 & 156 & 39.89 & 38.17 & 38.62 &  72 & 192 & 40.73 & -21.19\\
NGC~4085$^*$ & 17.0 &  43 & 39.56 & 38.24 & 38.43 &  64 & 143 & 40.13 & -19.83\\
NGC~4102$^*$ & 17.0 &  36 & 40.24 & 37.84 & 38.32 &  80 & 228& 40.94 & -21.91\\
NGC~4183 & 17.0 &   6 & 38.80 & 37.89 & 38.12 &  68 &  95 & 39.11 & -18.65\\
NGC~4389 & 17.0 &  21 & 39.44 & 38.32 & 38.70 &  74 & 151 & 40.11 & -19.18\\
NGC~6207 & 17.4 &  58 & 39.80 & 37.86 & 38.34 &  82 & 283 & 38.96 & -19.47\\
NGC~4701 & 20.5 &  48 & 39.47 & 37.98 & 38.30 &  80 & 249 & 39.50 & -19.52\\

\enddata

\tablecomments{(1): Name of galaxy. (2): Distance. (3): Number of
H\,{\sc ii} regions identified. (4), (5) and (6): H$\alpha$ luminosity of the
brightest, faintest and median H\,{\sc ii} region. (7) and 
(8): diameter of largest and median H\,{\sc ii} region. 
(9) and (10): H$\alpha$ luminosity in H\,{\sc ii}
regions and absolute F160W ($H$-band) 
magnitude over the central kpc.\\
$^*$ Nuclear emission excluded from H\,{\sc ii}
region statistics (columns (3) through (8)).\\
Galaxies excluded from
the statistical analysis of H\,{\sc ii} regions: 
NGC~247 (faint), NGC~891 (dust and edge-on)
NGC~2681 (nuclear emission only), 
NGC~2683 (dust and edge-on), NGC~2841 (nuclear emission only),
NGC~4293 (nuclear emission only and dust)
NGC~4395 (nuclear emission only), NGC~4571 (faint),
IC~4710 (nuclear emission only) 
and  NGC~6744 (nuclear emission only).} 
\end{deluxetable}

\section{Statistical properties} 

\subsection{Resolution effects} 
One of the most important issues when analyzing the properties of the
H{\sc ii} region LF is the role of observational parameters, especially
spatial resolution. Until now, most H{\sc ii} region  LFs were based
upon ground-based H$\alpha$ imaging with spatial resolutions ranging
from 0.8 to a few arcsec, or from 5\,pc (in M31; Walterbos \& Braun
1992)  to a few hundred pc (KEH; Rand 1992; Knapen et al. 1993; Banfi
et al. 1993; Rozas et al. 1996a, Knapen 1998), depending upon the
distance of the galaxy.  Rand (1992) discussed the effects of blending
on the properties of the H{\sc ii} region LF. Such blending, e.g., in
the case where a smaller H\,{\sc ii} region is spatially coincident with
a larger one and is not cataloged, is expected to occur more frequently
as the spatial resolution decreases. Rand (1992) concluded from his
modeling of this problem that blending does not significantly affect the
measured slope of the H\,{\sc ii} region LF as determined from his
ground-based image of M51.

More recently, Pleuss et al. (2000) specifically studied the impact of
different spatial resolution on, among other parameters, the diameter
distribution and LF of H\,{\sc ii} regions. They used {\it HST} archive
H$\alpha$ images of selected areas of the spiral galaxy M101, and
degraded these to a typical  ground-based resolution of 0.8
arcsec (FWHM).  The linear scales sampled by their high and low
resolution images were 3.6 and 77.6 pc/pixel, respectively. Although the
integral diameter distributions of the H\,{\sc ii} regions are
significantly different at high and low resolution (as expected, see
Knapen 1998), the LF slopes are only slightly different, being shallower
in the low resolution case. A similar effect is found by Scoville et
al. (2001)  who compared their {\it HST} (both WFPC2 and NICMOS)
H\,{\sc ii}  region LF with that of Rand (1992) for M51, with the
high-resolution LF significantly steeper than the low-resolution
one. Not only are the fitted slopes quite different 
but the methods used to arrive at them, so the comparison is less direct
than in the work by Pleuss et al. and Rand. It must be kept in mind that
the use of higher spatial resolution imaging also introduces some 
problems. For instance, H\,{\sc ii} regions may be
over-resolved, and individual components or stars within what ought to
be considered one single H\,{\sc ii} region may be cataloged as
separate, and by implication smaller and less luminous, ones.  

It is outside the scope of the current paper to discuss in detail the
problems of resolution on the statistical results on ensembles of
H\,{\sc ii} regions in spiral galaxies. However, we do briefly summarize
a few basic considerations which give some idea of physical
scales. Firstly, consider the ideal case of a Str\"omgren sphere, where
the radius of an H\,{\sc ii} region can be approximated in terms of the
number of ionizing photons and a number of factors which depend on the
electron temperature and density (see e.g., Str\"omgren 1939; Osterbrock
1989).  One finds that the radius of the Str\"omgren sphere of a main
sequence O5 star is 108\,pc, whereas that of a B0.5 star would be 12\,pc
(Osterbrock 1989). In sites of strong SF, massive stars tend to be
clustered in OB associations, so the expected sizes of the H\,{\sc ii}
regions ionized by these associations would be of the order of a few
hundred pc. Secondly, Walterbos \& Braun (1992) estimate an upper limit
for the H$\alpha$ luminosity of planetary nebulae (PN) in M31 of
$4\times 10^{35}\,{\rm erg}\,{\rm s}^{-1}$. For comparison, the
H$\alpha$ luminosity for H\,{\sc ii} regions ionized by an O5 star is
$6\times 10^{37}\,{\rm erg}\,{\rm s}^{-1}$ (this corresponds to the
approximate limit of an H\,{\sc ii} region ionized by a single star or
multiple stars), whereas for a B0.5 star it is $1\times 10^{35}\,{\rm
erg}\,{\rm s}^{-1}$.  The median H$\alpha$ luminosities of NGC~598 and
NGC~6207, the closest galaxies in our sample, are 4 and $5\times
10^{35}\,{\rm erg}\,{\rm s}^{-1}$, respectively, or close to the limit
for PN, and we have thus excluded these two galaxies from our
statistical analysis. For the remaining galaxies, the H$\alpha$
luminosities of the detected H\,{\sc ii} regions are always above the
limit of PN, so they are likely to correspond to real H\,{\sc ii}
regions.

\begin{figure*}
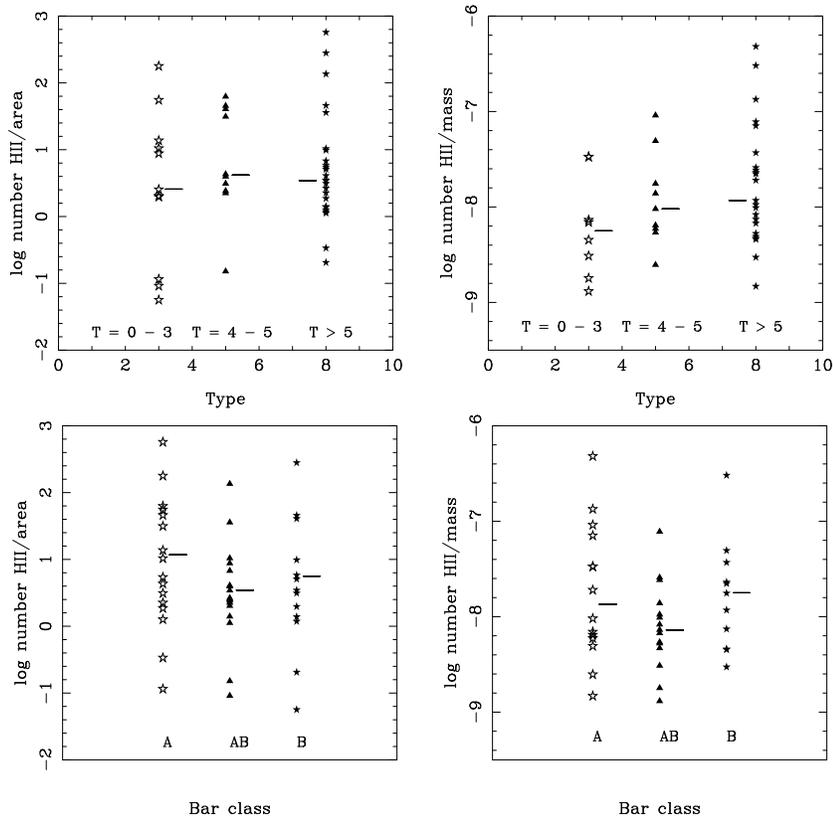

\figurenum{1}
\plotfiddle{figure1b.ps}{425pt}{-90}{50}{50}{-200}{410}
\plotfiddle{figure1a.ps}{425pt}{-90}{50}{50}{-200}{1000}
\vspace{-19.8cm}
\caption{{\it Upper left panel:} Distribution of the number of H\,{\sc
ii} regions per unit area for types $T = 0 -3$, $T=4-5$ and $T> 5$
(given in Table~1). {\it Upper right panel:} Same as left panel but with
number of H\,{\sc ii} regions per stellar mass unit (as traced by the
$H$-band luminosity). {\it Lower panels:} Same as upper panels but
comparing with the bar class. All diagrams are for galaxies with distances
$d \ge 2\,{\rm Mpc}$. The horizontal ticks next to the 
symbols show the median value of the distribution.}
\end{figure*}

\subsection{Number of H\,{\sc ii} regions} 

KEH analyzed the properties of disk H\,{\sc ii} regions of a sample of
nearby spiral and irregular galaxies. They found that the number of
H\,{\sc ii} regions (normalized to the total $B$-band luminosity)
increases by a factor of 6 from Hubble type Sb to Sc, and by another
50\% from Sc to irregulars.  KEH observed a similar behavior with the
number of H\,{\sc ii} regions per unit area, and no significant
differences between the H\,{\sc ii} regions in the inner and outer
halves of the disk of a given galaxy.

From Table~2, it can be seen that within a given distance bin, there is
a large dispersion in the number of H\,{\sc ii} regions detected in the
circumnuclear regions of galaxies. We can determine whether there is a
relation between the number of central H\,{\sc ii} regions per unit area
in kpc$^2$ (so the distance effect on the region covered by the images
can be attenuated) and the morphological type (we use the types $T$
given in Table~1) for all galaxies. The comparison is presented in the
upper left panel of Figure~1, where galaxies have been divided into
three morphological groups: $T = 0-3$, 
$T=4-5$ and $T>5$. We only include
galaxies with distances $d \ge 2\,{\rm Mpc}$, and plot
galaxies with only nuclear emission (see notes of Table~2) as having
one H\,{\sc ii} region detected. We also indicate the 
value of the median of the distribution.
This diagram does not
show the tendency for later-type galaxies to have more circumnuclear
H\,{\sc ii} regions per unit area than the earlier types, found by
KEH for H\,{\sc ii} regions in the disks of galaxies. 

In the upper right panel of Figure~1, we compare a related quantity,
that is the 
number of H\,{\sc ii} regions per unit near-infrared
$H$-band luminosity, which represents the stellar mass in the region
covered by the images. From this figure it can be seen that 
although the values of the median only rise slowly from early 
to late types, the
late-type galaxies show a tail in the distribution reaching higher 
numbers of circumnuclear H\,{\sc ii} regions per
unit stellar mass than earlier-types. This is caused by 
the fact that early
morphological type galaxies tend to have more luminous (and more
massive) bulges.  This in good agreement with the global behavior of
disk H\,{\sc ii} regions in nearby galaxies (KEH).

A similar comparison is carried out using the bar class (A, AB and B) to
determine whether the presence of a bar in a galaxy influences the
number of circumnuclear H\,{\sc ii} regions. 
The comparison is shown in the lower panels of Figure~1 for
the number of H\,{\sc ii} regions per unit area, and per unit stellar
mass.  There is no  tendency for barred galaxies to show more
H\,{\sc ii} regions per unit stellar mass or unit area.

\begin{figure*}
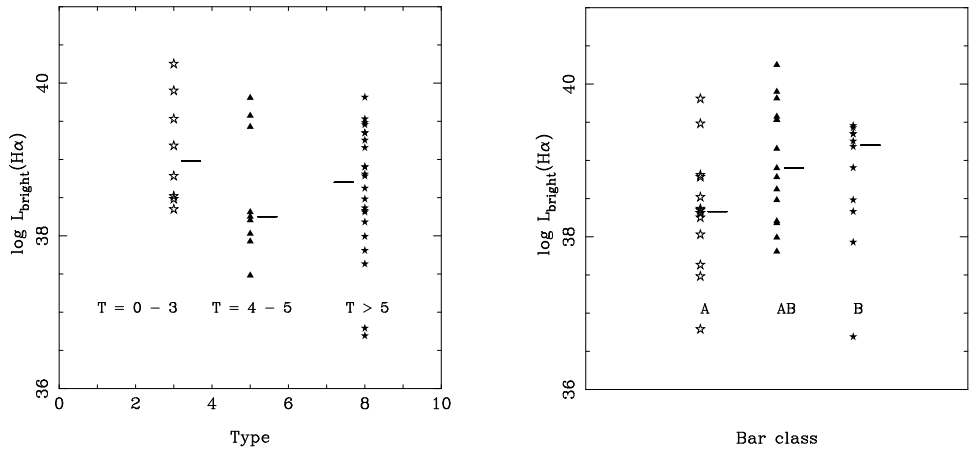

\figurenum{2}
\plotfiddle{figure2a.ps}{425pt}{0}{50}{50}{-200}{200}
\plotfiddle{figure2b.ps}{425pt}{0}{50}{50}{0}{635}
\vspace{-23cm}
\caption{H$\alpha$ luminosity of the brightest H\,{\sc ii} region as a 
function of the morphological type (left panel) 
and  bar class (right panel) for those 
galaxies in our sample with distances $d \ge 2\,{\rm Mpc}$.}
\end{figure*}

\subsection{H\,{\sc ii} region luminosities}
	
The accepted view is that the global SF activity in galaxies decreases
systematically from late-type to early-type spirals, a conclusion mainly
based on the integrated equivalent widths of H$\alpha$ (e.g., Kennicutt
\& Kent 1983 and Kennicutt 1998a for a review). This view is also
supported by the fact that the luminosities of the brightest H\,{\sc ii}
regions in the disks of early-type spiral galaxies are on average lower
than those of later types (KEH).  Moreover, Caldwell et al. (1991)
reported that in Sa galaxies there are no H\,{\sc ii} regions with
H$\alpha$ luminosities greater than $10^{39}\,{\rm erg}\,{\rm s}^{-1}$,
although this view has been challenged by Devereux \& Hameed (1997) and
Hameed \& Devereux (1999, and references therein). 
The SF activity of galaxy nuclei differs from the global behavior as found
by Ho, Filippenko, \& Sargent (1997a) who from an optical spectroscopic 
survey showed that 
early-type (S0--Sbc) galaxies have nuclei with  
higher H$\alpha$ luminosities  than late-type (Sc--I0) galaxies, 
with an increase in the median of about a factor of 9, for 
H\,{\sc ii} (starburst) type galaxies.

\begin{figure*}
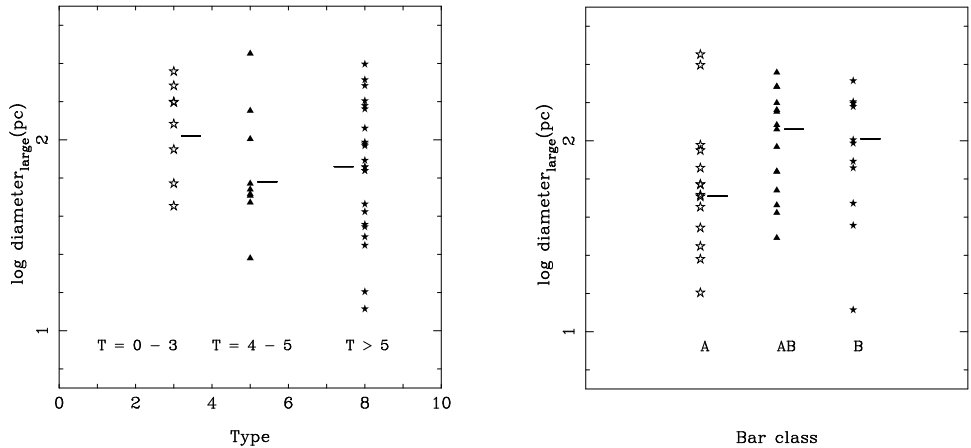

\figurenum{3}
\plotfiddle{figure3a.ps}{425pt}{0}{50}{50}{-200}{200}
\plotfiddle{figure3b.ps}{425pt}{0}{50}{50}{0}{635}
\vspace{-23cm}
\caption{As Figure~2, but for the diameter of the largest H\,{\sc ii}
region.}
\end{figure*}

When we compare the median luminosities of the circumnuclear H\,{\sc ii}
regions with the morphological type for galaxies within distance bins
(see Table~2), we find no clear evidence for later-type galaxies to show
brighter circumnuclear H\,{\sc ii} regions. For instance, within the
$d=17\,$Mpc distance bin, the median H$\alpha$ luminosity of H\,{\sc
ii} regions in the early-type spirals IC~750 and NGC~4192 are above the
average value within this bin (although we note that IC~750 is an
interacting galaxy). A similar behavior is found within the 4--5\,Mpc
distance bin, where the early-type galaxies NGC~3593 and NGC~4826 host
relatively bright median H\,{\sc ii} regions. Although our number
of early-type spirals (that is, Sa and Sab) is small, 
there is no clear trend for earlier-type galaxies to
show fainter circumnuclear H\,{\sc ii} regions.

The luminosity of the brightest circumnuclear H\,{\sc ii} region can be
used as an estimate of the age of the SF process (at least if the SF
occurred in an instantaneous burst), although this quantity is also
dependent upon the maximum number of ionizing stars per cluster, and
upon the upper mass cutoff of the IMF (see Feinstein 1997; Oey \& Clark
1998). An advantage of using the luminosity of the brightest H\,{\sc ii}
region over the mean luminosity is that is not sensitive
to the minimum size imposed for the detection of an H\,{\sc ii}
region, although the blending effects will be more important for the
more distant galaxies.

The left panel of Figure~2 shows the logarithm of the
H$\alpha$ luminosity of the brightest circumnuclear H\,{\sc ii} region
as a function of morphological type (grouped as in Figure~1) for
galaxies with distances $d> 2\,{\rm Mpc}$. For a given group of
morphological types, there is a spread of approximately two orders of
magnitude in the luminosity of the brightest H\,{\sc ii} region, where
part of the scatter is probably caused by blending effects in the more
distant galaxies. Although the median values of the H$\alpha$ luminosity for 
the brightest H\,{\sc ii} region are similar for the early-types ($T=0-3$)
and late type ($T>5$), the distribution for the late-type
galaxies shows a tail reaching fainter luminosities. 
As the early-type galaxies are biased towards more distant galaxies, 
we checked that the low luminosity tail in the late types 
is not caused by the 
distance effect  by examining the distribution
excluding galaxies with distances  $d \ge 12\,$Mpc. 
Again we find  no tendency for the late type ($T > 5$) galaxies to show
more luminous brightest H\,{\sc ii} regions than their 
early type counterparts.

This result is in clear contrast with the behavior of disk H\,{\sc ii}
regions. Kennicutt (1988) found that the brightest H\,{\sc ii} regions
in irregular galaxies are approximately 6 times brighter than those in
Sbc-Sc galaxies of the same absolute blue magnitude, and approximately
50 times brighter than their counterparts in Sab--Sb galaxies. This
result is always referred to disk H\,{\sc ii} regions. Moreover, 
three of
the six galaxies in our sample with Hubble morphological types S0/a to
Sab show $L_{\rm br}({\rm H}\alpha)$ near or above the limit given
in Caldwell et al. (1991), whereas two only show faint central
emission. Although the numbers are small, it is evident that some
early-type galaxies can harbor very luminous central H\,{\sc ii}
regions.  Possible explanations for these discrepant results are
resolution effects, reduced extinction in our Pa$\alpha$ data, but more
likely a different behavior of circumnuclear H\,{\sc ii} regions versus
disk H\,{\sc ii} regions.

The bar potential in a disk galaxy, which by its
non-axisymmetric nature can funnel gaseous material from the disk
inwards toward the central regions, is usually invoked 
as a triggering of the SF in the centers of galaxies 
(e.g., Shlosman, Begelman \& Frank
1990).  There, the gas can accumulate, possibly near one or more inner
Lindblad resonances, become gravitationally unstable, and this can lead to
enhanced massive star formation. 

We can now check this scenario by comparing 
the luminosity of the brightest H\,{\sc ii} region with the bar class 
(right panel of Figure~2). There is a clear tendency for
barred galaxies to show more luminous brightest H\,{\sc ii} regions than
their unbarred counterparts, as can be seen from the increasing 
value of the median of the brightest H\,{\sc ii} region distribution 
from unbarred (A) to strongly barred (B) galaxies. 
Although the median distance of the strongly barred  galaxies is 
higher than that of unbarred galaxies, the distance effect
alone is not 
enough to account for the differing medians of the distributions of the 
brightest H\,{\sc ii} regions. If galaxies in the 
Virgo cluster, which populate 
40\% of the strongly barred galaxy bin and  50\% of the
weakly barred galaxy bin, are excluded, we find that the median of 
the distribution
of the brightest H\,{\sc ii} regions in barred galaxies (AB+B)
is slightly higher than in unbarred galaxies, but on average 
the brightest H\,{\sc ii} regions in barred galaxies are
a factor of five brighter than those in unbarred
systems. A K-S test shows that there is
only a probability of $P_{\rm K-S}=0.11$ that both distributions
are drawn from the same  population, for galaxies with 
$d<12\,$Mpc. 
The scenario described above is proven here
in a statistical and qualitative way, a conclusion that fits in well
with the long known preference for central starburst galaxies to be
barred (e.g., Heckman 1980; Balzano 1983; Devereux 1987), and with the
occurrence of circumnuclear ring-like regions of much enhanced SF in
barred galaxies (Knapen 1999).

\subsection{H\,{\sc ii} region sizes}

The size of an H\,{\sc ii} region is not as
easy to measure as its luminosity, because the former quantity
is observationally mostly influenced by pixels at the perimeter of the
H\,{\sc ii} region, thus maximizing the sensitivity to blending and
background noise, whereas the luminosity is determined mostly by the
bright pixels near the centers of H\,{\sc ii} regions. Nevertheless, Hodge
(1986) found that late-type galaxies (Sc and Irr) tend to have larger
H\,{\sc ii} regions than early-type galaxies.  Kennicutt (1988) measured
the sizes of the first ranked H\,{\sc ii} regions in a sample of spiral
and irregular galaxies, and reached a similar conclusion, although the
trend was not as significant as for the luminosity of 
the brightest H\,{\sc ii}
regions.  

In Figure~3 (see also Table~2) we compare the size of the
largest H\,{\sc ii} region (which is not necessarily  the brightest
one) with the morphological type and bar class. As found with the
luminosity of the brightest H\,{\sc ii} region, there is no clear trend
with the morphological type. The size of the largest H\,{\sc ii} regions 
in barred galaxies is on average larger  than 
for unbarred galaxies. As we did for the 
luminosity of the brightest H\,{\sc ii} region, 
if the distribution is examined only considering galaxies with distances
$d<12\,$Mpc (to avoid distance biases), the trend is maintained with a 
probability of only $P_{\rm K-S}=0.16$ that 
the size distributions of the largest H\,{\sc ii} region in barred 
and unbarred galaxies are drawn from the same population.

In summary, we can state that while the physical properties (luminosity
and size) of the circumnuclear H\,{\sc ii} regions are less sensitive to
the morphological type of their host galaxy than those of disk H\,{\sc ii}
regions, the presence of a bar produces on average larger and more
luminous circumnuclear H\,{\sc ii} regions than in non-barred galaxies.

\subsection{Central kpc H$\alpha$ luminosities and specific SFRs}

The use of narrow band Pa$\alpha$ imaging, as opposed to 
spectroscopy, allows to 
evaluate the strength of the SF activity in the centers of galaxies, 
covering the same physical size for all galaxies. For this purpose,  
we use two quantities, the total H$\alpha$ emission (only from H\,{\sc ii}
regions) and the SFR per stellar mass unit (also
known as specific SFR).  The SFR per unit
stellar mass can also be understood 
as an integrated pseudo equivalent width  of Pa$\alpha$, since both are 
basically the result of dividing the Pa$\alpha$ flux 
by the nearby $H$-band continuum. Thus, the specific SFR or the 
pseudo equivalent width can be  used as an estimate of the efficiency of the 
SF processes with respect to the stellar mass, the age of the SF process,
or most likely, a combination of both. 

We evaluated the H$\alpha$ luminosity and specific SFR for the central 
kpc of each galaxy in our sample except for those with distances
of less than 4\,Mpc as the NIC3 field of view does not
cover the central kpc. The H$\alpha$ emission was calculated by adding  
the luminosity of all the H\,{\sc ii} regions whose centers are 
within the central circular kpc area. We  performed aperture photometry 
(circular apertures) covering the central kpc
on the $H$-band images, and used the mass-to-light
ratio given in Thronson \& Greenhouse (1988) to estimate the central stellar 
masses. The H$\alpha$ luminosities and absolute $H$-band magnitudes over
the central kpc for the galaxies in our sample are listed in the 
last two columns of Table~2.
The SFRs are computed using the relation provided 
in Kennicutt (1998b). Note that these quantities are obviously independent
of the distance of the galaxy, and resolution effects. 

The comparison of the central H$\alpha$ luminosity 
with the morphological type is presented in the left panel of Figure~4,
which clearly shows enhanced SF in the central kpc of early type galaxies.
This confirms the tendency found by B\"oker
et al. (1999) of higher average  surface brightnesses
(over the field of view of the
images)  for earlier morphological types, as well as results by Ho 
et al. (1997a,b)  from  optical spectroscopy. However, 
when the specific SFRs are plotted versus the morphological 
type there is no
clear increase of the specific SFRs for earlier type galaxies, mainly
because these galaxies tend to have more massive bulges (Figure~5, left
panel).

The effect of a bar on both the central kpc H$\alpha$ 
luminosity and specific SFRs is evaluated in the right panels of
Figure~4 and 5.  The distributions of the central kpc H$\alpha$
luminosities (only in H\,{\sc ii} regions) show
slightly increasing values of the median from unbarred to barred galaxies.
However, as also found 
by Ho et al. (1997b) using optical spectroscopy 
(although their spectroscopy did not cover the same physical size
for all galaxies), the distribution for barred (AB+B) galaxies has
a tail reaching higher central H$\alpha$ luminosities ($\log L({\rm H}\alpha)
_{\rm kpc}> 40\,{\rm erg\,s}^{-1}$) 
than that of unbarred galaxies. This result is not
of a high statistical significance, with a probability that the distributions
of central H$\alpha$ luminosity of barred and unbarred 
galaxies are drawn from the same population of $P_{\rm K-S}=0.33$.
Ho et al. (1997b) 
noted that the trend for barred galaxies to show enhanced central
SF was more prominent if only early-type (S0/a--Sbc) were considered.
We reexamined their result but with all galaxies 
covering the same physical 
size (the central kpc). Although this subsequent subdivision 
of our sample implies small number statistics, we find 
significant differences, with early type ($T<5$) unbarred galaxies showing 
a median of 
$\log L({\rm H}\alpha)_{\rm kpc} = 39.3\,{\rm erg\,s}^{-1}$
(7 galaxies) in contrast with their barred counterparts:   
$\log L({\rm H}\alpha)_{\rm kpc} = 40.1\,{\rm erg\,s}^{-1}$
(9 galaxies). On the other hand, the 
late type galaxies, both barred and unbarred, 
show similar values of the median 
of the distributions:
$\log L({\rm H}\alpha)_{\rm kpc} = 39.2\,{\rm erg\,s}^{-1}$.
Clearly, the luminosity enhancement in the central kpc of
barred galaxies occurs almost entirely in early type galaxies.

We also find enhanced specific SFRs (or pseudo equivalent 
widths of Pa$\alpha$) in the central kpc of 
barred galaxies  as seen
from the increasing median of the distributions
from unbarred to barred galaxies (see right panel of Figure~5). 
This is in good agreement with findings by 
Ho et al. (1997b) who used the equivalent width of H$\alpha$.
This provides further evidence in support 
of the expected enhancement of SF (both in terms of the efficiency 
and the younger ages) in
the centers of barred with respect to non-barred galaxies, due to gas
losing angular momentum in a bar, and falling inward from the disk to the
central region.

\begin{figure*}
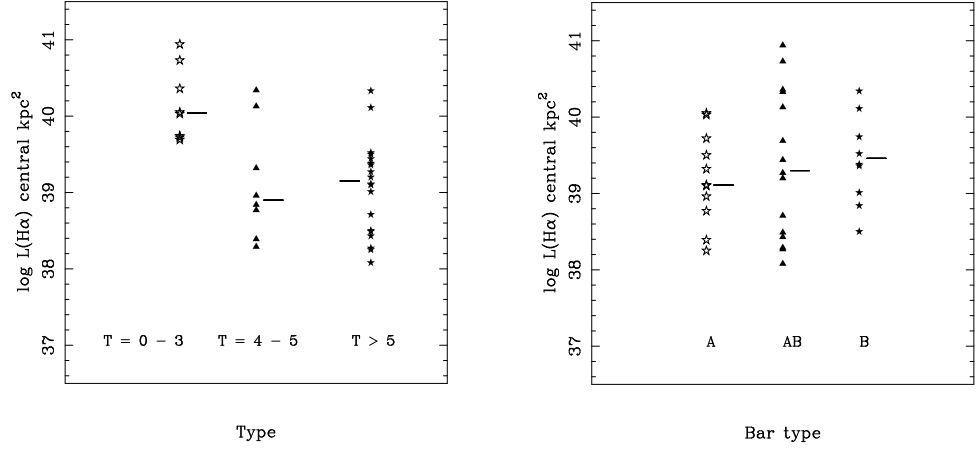

\figurenum{4}
\plotfiddle{figure4a.ps}{425pt}{0}{50}{50}{-200}{200}
\plotfiddle{figure4b.ps}{425pt}{0}{50}{50}{0}{635}
\vspace{-23cm}
\caption{As Figure~2, but for the H$\alpha$ luminosity 
in H\,{\sc ii} regions over the central kpc. }
\end{figure*}

\begin{figure*}
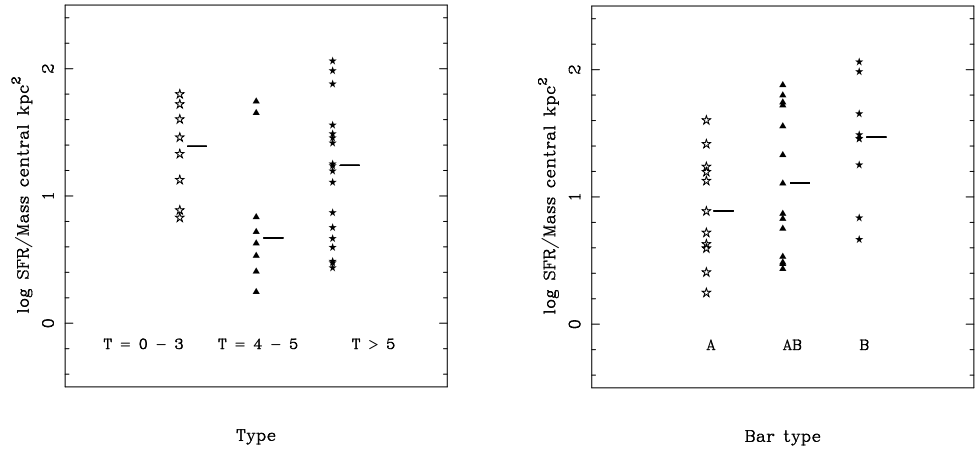

\figurenum{5}
\plotfiddle{figure5a.ps}{425pt}{0}{50}{50}{-200}{200}
\plotfiddle{figure5b.ps}{425pt}{0}{50}{50}{0}{635}
\vspace{-23cm}
\caption{As Figure~2, but for the SFR per 
unit stellar mass over the central kpc. }
\end{figure*}

\section{LFs and diameter distributions}

In this section we concentrate on those galaxies for which we cataloged
100 or more H\,{\sc ii} regions from the Pa$\alpha$ images of the
circumnuclear regions, and for which we can construct reliable LFs and
integral diameter distributions.

\begin{deluxetable}{llcc}
\tablewidth{10cm}
\tablecaption{Parameters of the fitted H\,{\sc ii} region LF for selected
galaxies.}
\tablehead{\colhead{Galaxy}  & \colhead{Type} & 
\colhead{$\alpha$} & \colhead{$\log L({\rm H}\alpha)$} \\
        &      &          & range (erg s$^{-1})$}
\startdata

NGC~2903 & SB(s)d      & $-1.65 \pm 0.16$ & $37.5-39.5$\\
NGC~3077 & I0pec       & $-1.70 \pm 0.07$ & $36.7-38.3$\\
IC~750   & Sab         & $-2.02 \pm 0.07$ & $38.3-39.9$\\
NGC~3593 & SA(s)0/a    & $-1.68 \pm 0.16$ & $37.4-38.8$\\
NGC~4449 & IBm         & $-1.85 \pm 0.05$ & $36.7-38.5$\\
NGC~4826 & (R)SA(rs)ab & $-2.27 \pm 0.16$ & $36.9-38.5$\\
NGC~5055 & SA(rs)bc    & $-2.32 \pm 0.18$ & $37.3-38.3$\\
NGC~6946 & SAB(rs)cd   & $-1.83 \pm 0.05$ & $36.9-39.5$\\
\enddata
\end{deluxetable}

\begin{figure*}
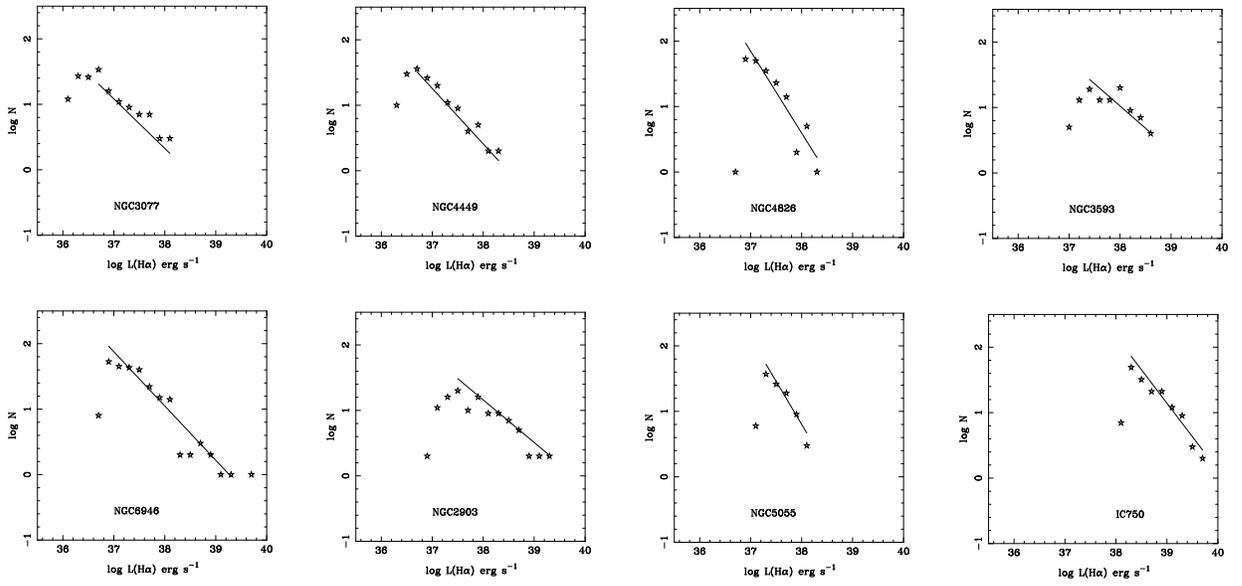

\figurenum{6}
\plotfiddle{figure6a.ps}{425pt}{0}{30}{30}{-240}{200}
\plotfiddle{figure6b.ps}{425pt}{0}{30}{30}{-120}{635}
\plotfiddle{figure6c.ps}{425pt}{0}{30}{30}{0}{1070}
\plotfiddle{figure6d.ps}{425pt}{0}{30}{30}{120}{1505}
\vspace{4cm}
\plotfiddle{figure6e.ps}{425pt}{0}{30}{30}{-240}{1940}
\plotfiddle{figure6f.ps}{425pt}{0}{30}{30}{-120}{2375}
\plotfiddle{figure6g.ps}{425pt}{0}{30}{30}{0}{2810}
\plotfiddle{figure6h.ps}{425pt}{0}{30}{30}{120}{3245}
\vspace{-114cm}
\caption{Luminosity Functions of galaxies with 100 or more H\,{\sc ii}
regions, and power law fits.}
\end{figure*}

\subsection{H\,{\sc ii} region  LFs} 

The differential LFs of circumnuclear H\,{\sc ii} regions for the
selected galaxies are presented in Figure~6, as log-log plots in
luminosity bins of 0.20\,dex. We also show the best fit to the slope
(with $\alpha = {\rm slope} - 1$) in the figures.  The parameters of the
fit to the circumnuclear H\,{\sc ii} region LF ($\alpha$ and the
luminosity range used for fitting the slope) are given in Table~3. The
slopes of the LFs are very much within the range found for LFs for disks
of spiral galaxies from ground-based imaging, perhaps surprisingly so
given the very different data sets used (Pa$\alpha$ vs. H$\alpha$
imaging, high vs.  relatively low spatial resolution, circumnuclear
vs. disk H\,{\sc ii} regions). 
This similarity may well indicate common
physical processes underlying the massive SF in circumnuclear and disk
areas in spiral galaxies.

The peak of the H\,{\sc ii} region LF, which corresponds to the lower
luminosity end where the LF is complete, is relatively constant in these
galaxies, and occurs at approximately $\log L({\rm H} 
\alpha)= 37 \,{\rm erg}\,{\rm
s}^{-1}$, a value similar to that found by Scoville et al. (2001) in
M51. The one exception is IC~750, a galaxy which is not only
interacting, but also by far the most distant of the galaxies discussed
in this section.  We find that the LFs of the circumnuclear H\,{\sc ii}
regions extend to H$\alpha$ luminosities of the order of
$\log L({\rm H}\alpha) = 
38.3-38.8\,{\rm erg}\,{\rm s}^{-1}$ in most cases, but the
LFs cover luminosities of up to $\log L({\rm H}\alpha)={39.9}\,{\rm erg}\,{\rm s}^{-1}$
in those galaxies with enhanced SF (NGC~2903, NGC~6946 and IC~750).
Ground-based LFs for disk H\,{\sc ii}
regions usually extend out to $\log L({\rm H}\alpha) 
\simeq {40}\,{\rm erg}\,{\rm
s}^{-1}$ (e.g., KEH; Rozas et al.  1998a), although the
high luminosity end depends upon the galaxy distance, as it is 
the case of our sample.

\begin{figure*}
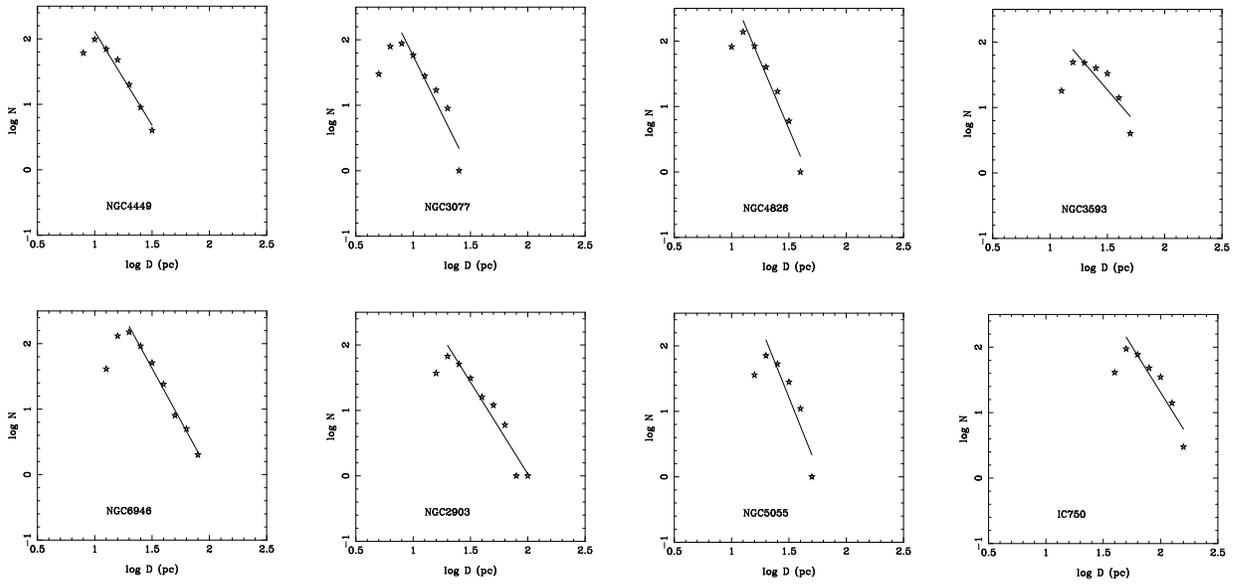

\figurenum{7}
\plotfiddle{figure7a.ps}{425pt}{0}{30}{30}{-240}{200}
\plotfiddle{figure7b.ps}{425pt}{0}{30}{30}{-120}{635}
\plotfiddle{figure7c.ps}{425pt}{0}{30}{30}{0}{1070}
\plotfiddle{figure7d.ps}{425pt}{0}{30}{30}{120}{1505}
\vspace{4cm}
\plotfiddle{figure7e.ps}{425pt}{0}{30}{30}{-240}{1940}
\plotfiddle{figure7f.ps}{425pt}{0}{30}{30}{-120}{2375}
\plotfiddle{figure7g.ps}{425pt}{0}{30}{30}{0}{2810}
\plotfiddle{figure7h.ps}{425pt}{0}{30}{30}{120}{3245}
\vspace{-114cm}
\caption{Differential diameter distributions of galaxies with 100 or
more H\,{\sc ii} regions, and power law fits.}
\end{figure*}

Although our numbers are small, we point out another
effect, namely that the slopes of the circumnuclear H\,{\sc ii}
region LFs tend to be shallower in later type galaxies, in good
agreement with findings by KEH and Banfi et al. (1993) for disk H\,{\sc
ii} regions. The only discrepant galaxy is NGC~3593.
Oey \& Clarke (1998) interpreted the steeper slopes in
early type galaxies as due to a decrease in the upper limit to the
number of stellar clusters. They based their argument on
the claim by Caldwell et al. (1991) that Sa galaxies only show maximum
H$\alpha$ luminosities in the range 
$\log L{\rm (H}\alpha) = 38.0-38.5\,$erg s$^{-1}$,
and that they may thus only contain unsaturated clusters (clusters with
poor stellar statistics). We have already shown above 
that that may not be the case for all early type galaxies. Other
possibilities for the steeper LF slopes in early type galaxies would be
a change in the slope of the stellar cluster distribution, or an aging
effect.

In a series of studies of the properties of H\,{\sc ii} regions of grand
design spiral galaxies, Beckman et al. (2000, and references therein)
have pointed out the presence of a glitch in the LF at a luminosity of
$\log L({\rm H}\alpha) = 38.6 \pm 0.1\,{\rm erg}\, {\rm s}^{-1}$,
accompanied by a steepening of the slope of the LF for luminosities
above this value.  Beckman et al.  interpreted this luminosity 
and other pieces of observational evidence to argue this 
luminosity marks the 
transition between ionization-bounded and density-bounded H\,{\sc ii}
regions. 
This interpretation has, however, been disputed by Pleuss et al. (2000),
who claim that the glitch in the LF, as well as some of the other
observational hints, may be artifacts due to blending of H\,{\sc ii}
regions as a direct result of the limited, ground-based, spatial
resolution of the data used by Beckman et al. (2000).  This is a debate
that is outside the scope of this paper, but we do point out that there
is no clear evidence for the presence of a glitch or change of slope
near the luminosity reported by Beckman et al. (2000) from the
circumnuclear H\,{\sc ii} region LFs in Figure~6. From our few 
LFs we do not find a change of slope at $\log L_({\rm H}\alpha) =
38.6 \pm 0.1\,{\rm erg}\,{\rm s}^{-1}$, presumable because with our
{\it HST} imaging we resolve luminous H\,{\sc ii} regions into several
less luminous ones. Scoville et al. (2001) report similar results from
their {\it HST} study of H\,{\sc ii} regions in M51. 


\subsection{Integral diameter distributions}

The integral diameter distribution of H\,{\sc ii} regions in galaxies is
usually described as an exponential relation, 

\begin{equation} 
N(>D) = N_0{\rm e}^{-D/D_0}, 
\end{equation}

\noindent which thus relates the number $N$ of H\,{\sc ii} regions with
diameters larger than $D$ to the diameter $D$, where $D_0$ is the
characteristic size of the distribution (van den Bergh 1981; Hodge 1983;
Rozas, Knapen \& Beckman 1996b).  However, as discussed by Kennicutt \&
Hodge (1986) for the case of ionization-bounded H\,{\sc ii} regions
within a neutral medium of uniform density, one would expect the
Str\"omgren radii to exhibit a power law distribution:

\begin{equation}
{\rm d}N(R) =  B\,R^{2-3\alpha}\,{\rm d}R = C\,R^{\beta}\,{\rm d}R.
\end{equation}

Indeed, recently Pleuss et al. (2000), using high resolution ({\it HST})
data, have shown that the size distribution of H\,{\sc ii} regions in
selected regions in M101 is better fitted with a power law than with an
exponential relation. Moreover, these authors suggest that some of the
previously fitted exponential forms to the integral diameter
distributions might have been caused by spatial resolution effects.

Since the resolution of our data is comparable to that of Pleuss et
al. (2000), we can address the question of whether the diameter
distributions for those galaxies with 100 or more H\,{\sc ii} regions
can be fitted with power laws. The distributions are presented in
Figure~7, and the power law fits are given in Table~4.  We confirm the
result reported by Pleuss et al. (2000) that at high spatial resolutions
the diameter distributions can be fitted as power laws, as expected for
ionization-bounded H\,{\sc ii} regions. Also listed in Table~4 are the
predicted values for the slope of the relation ($\beta_{\rm pred} = 2 -
3\,\alpha$, with the errors in $\beta_{\rm pred}$ computed as $\Delta
\beta_{\rm pred} = 3\,\Delta \alpha$). From this table it can be seen
that there is a relatively good agreement (to within the errors) 
between the fitted
values of $\beta$ and those predicted from the LF  under the above
assumption of a constant density. This good agreement is reflected in
the well behaved luminosity-volume relations (not shown). 
We also performed exponential fits to the diameter 
distributions, and found that generally the power law and 
the exponential forms provided fits to diameters distributions 
of similar statistical significance.
A caveat we
must mention here, though, is the measurement of 
diameters for H\,{\sc ii} regions in the
rather crowded circumnuclear areas under investigation in this paper is
not straightforward.

\begin{deluxetable}{lccc}
\tablewidth{10cm}
\tablecaption{Parameters of the fitted size 
distributions of H\,{\sc ii} regions in selected galaxies.}
\tablehead{\colhead{Galaxy}  & \colhead{$\beta$} & 
\colhead{$\log D$} & \colhead{$\beta_{\rm pred}$}\\
        &          & range (pc)}

\startdata
NGC~2903 & $-3.81 \pm 0.24$ & $1.3-2.0$ & $-2.95 \pm 0.18$\\
NGC~3077 & $-4.53 \pm 0.59$ & $0.9-1.4$ & $-3.10 \pm 0.21$\\
IC~750   & $-3.81 \pm 0.48$ & $1.7-2.4$ & $-4.06 \pm 0.21$\\
NGC~3593 & $-3.04 \pm 0.48$ & $1.2-1.7$ & $-3.04 \pm 0.48$\\
NGC~4449 & $-3.86 \pm 0.21$ & $1.0-1.5$ & $-3.55 \pm 0.12$\\
NGC~4826 & $-5.15 \pm 0.44$ & $1.1-1.6$ & $-4.75 \pm 0.48$\\
NGC~5055 & $-5.38 \pm 0.87$ & $1.3-1.7$ & $-4.96 \pm 0.54$\\
NGC~6946 & $-4.11 \pm 0.15$ & $1.3-1.9$ & $-3.49 \pm 0.15$\\
\enddata
\end{deluxetable}

\section{Concluding remarks}

We have used {\it HST}/NICMOS Pa$\alpha$ observations of a sample of
nearby galaxies to analyze for the first time the properties of the
circumnuclear H\,{\sc ii} regions at spatial resolutions ranging from 1
to 30\,pc. Our sample was selected from the {\it HST}/NICMOS snapshot
survey of nearby galaxies (in $H$ and Pa$\alpha$) by B\"oker et
al. (1999).  It 
encompasses 52 galaxies with morphological types
S0/a and later, and velocities $v \le 1000\,{\rm
km}\,{\rm s}^{-1}$. The high spatial resolution of these observations overcomes
some of the blending problems encountered by previous ground-based
studies, and
allows a detailed comparison of the properties of circumnuclear and disk
H\,{\sc ii} regions. Kennicutt (1998a) provides an excellent review on
the subject of the SF properties in galaxies, and the dependencies on
different factors such as the Hubble morphological type, presence of
bars, interactions, and gas content, for both disks and circumnuclear
regions. However, most previous works on circumnuclear SF
properties were based on integrated properties rather than on the study
of the properties of the individual H\,{\sc ii} regions.

We briefly summarize our main results, classed in three categories, below.

\subsection{Relation to morphological type}

We find that the physical properties of individual circumnuclear H\,{\sc
ii} regions, as represented by typical H\,{\sc ii} regions (median
H$\alpha$ luminosity and size) and by first ranked H\,{\sc ii} regions
(the most luminous and the largest H\,{\sc ii} regions), are not
strongly dependent on the morphological type of the host galaxy. 
Also, we find no
relationship between the number of circumnuclear H\,{\sc ii} regions per
unit area and the morphological type. The behavior of the physical
properties of circumnuclear H\,{\sc ii} regions is in clear contrast
with that of disk H\,{\sc ii} regions which tend to be larger, brighter and
more numerous (per unit area) for late type spiral and irregular
galaxies (Kennicutt 1988; KEH).

Whereas the morphological type does not seem to be one of the dominant
factors in determining the physical properties of the individual
circumnuclear H\,{\sc ii} regions, it does influence the global
circumnuclear SF properties. The H$\alpha$ 
luminosities over the central kpc (only based upon emission
from H\,{\sc ii} regions) are significantly enhanced in early-type
(S0/a--Sb)
galaxies when compared with late-type galaxies. This confirms the findings of
B\"oker et al. (1999) for the average central surface 
Pa$\alpha$ brightnesses (H\,{\sc ii} region and diffuse
emission) over the field of view of the images, and other 
studies based upon optical spectroscopy (e.g., Ho et al. 1997a,b and 
references therein). When the SFR per
unit stellar mass are compared with the morphological type, the trend
disappears because it is offset by a trend in the opposite direction,
where earlier type galaxies tend to have more massive bulges than later
type galaxies.

\subsection{Relation to bars}
A relation between the presence of a bar and the properties of the
circumnuclear SF is expected since bars are predicted to provide an
efficient mechanism to transport gaseous material from the disks of
galaxies into the central regions, and as a consequence bars may trigger the
SF in the circumnuclear regions of galaxies. Indeed, barred
galaxies have enhanced H$\alpha$ luminosities (mostly occurring 
in early type galaxies) and SFR per
unit stellar mass (or integrated pseudo equivalent
width of Pa$\alpha$) over the central 1 kpc area when compared
to unbarred galaxies. Whereas
the number of H\,{\sc ii} regions per area or per unit stellar mass are
not enhanced in barred with respect to non-barred galaxies, the
first-ranked H\,{\sc ii} regions (both in terms of diameter and
luminosity) are on average more luminous and larger in barred than 
in unbarred galaxies. This in conjuction with 
results from many other works provide further evidence 
that bars are efficient in triggering the SF 
in the central regions of galaxies, although, 
as pointed out among others by Ho et al. (1997b), the presence of a bar is 
neither  a necessary 
nor a sufficient condition for SF to occur. 

\subsection{LFs and diameter distributions}

We have analyzed the LFs and integral diameter distributions of the eight
galaxies in our sample with one hundred or more circumnuclear H\,{\sc
ii} regions. The LFs of the circumnuclear H\,{\sc ii} regions 
extend to H$\alpha$ luminosities of $\log L({\rm H}\alpha) = 
38.3-38.8\,{\rm erg}\,{\rm
s}^{-1}$, whereas in galaxies with enhanced SF the LFs
reach $\log L({\rm H}\alpha) = 39.7\,{\rm erg}\,{\rm
s}^{-1}$. We fitted power 
law slopes of the circumnuclear H\,{\sc ii} region LFs, of between 
$\alpha=-2.3$ and $\alpha = -1.7$, values 
which  are exactly within the range of
slopes reported for spiral disks  from ground-based
observations (e.g., KEH).  This suggests that the physical processes
determining the massive SF in disks and circumnuclear regions must be
common.
Although we can only fit slopes to the H\,{\sc ii} region LFs for a very
limited number of galaxies, we do confirm the relation between LF slope
and galaxy type, with late type galaxies showing shallower
LFs.

The integral diameter distributions for these eight galaxies 
are fitted with power laws whose indices are in a relatively 
good agreement with those predicted from the LF, assuming 
that the H\,{\sc ii} regions are ionization-bounded within 
a medium of uniform density. We find, however, that
the more commonly used exponential form provides fits of similar
statistical significance.

\subsection{Final comments}

With the detailed analysis of narrow band H$\alpha$ and Pa$\alpha$
images of spiral galaxies obtained with the {\it HST}, we have
recently entered a new era in research on extragalactic H\,{\sc ii}
regions (Pleuss et al. 2000; Scoville et al. 2001; this paper). One of
the main conclusions from all these works is that although blending
may affect results based on ground-based
narrow-band imaging, one of the main parameters describing statistical
ensembles of H\,{\sc ii} regions from {\it HST} data, 
namely the slope of the LF, is still
within the previously observed values. The recent 
conclusions emerging 
from studies using {\it HST} imaging, which are based either on partial
imaging of the disk of one individual galaxy, or, as in our case, on
imaging of the circumnuclear parts only of a sample of spiral galaxies,
will need to be confirmed by more extensive studies using
high-resolution narrow-band imaging, but contain
important clues to the underlying physical processes of massive SF in
galaxies.

\section*{Acknowledgments}

We are grateful to an anonymous referee for useful comments that helped
improve the paper.
It is a pleasure to thank  Dr. C. H. Heller for providing us with the {\sc
region} software, used to analyze the properties of the H\,{\sc ii}
regions, and Drs. J.E. Beckman and A. Zurita for comments on an
earlier version of the manuscript.

This research has made use of the NASA/IPAC Extragalactic Database (NED) 
which is operated by the Jet Propulsion Laboratory, California Institute 
of Technology, under contract with the National
Aeronautics and Space Administration.

\end{document}